\documentclass[amsmath]{iopart}
\usepackage[left=2cm,right=2cm]{geometry}
\usepackage{graphicx} % include figure files
\usepackage{bm} %bold math
\usepackage{longtable}
\usepackage{multirow}
\usepackage{enumerate}
\usepackage{amssymb}

\newcommand{\bra}[1]{\langle #1 |}
\newcommand{\ket}[1]{| #1 \rangle}

\newcommand{\crea}[2]{\hat #1^{\dagger}_{#2}}
\newcommand{\anni}[2]{\hat #1_{#2}}

\begin{document}
 
\title{Infrared-dressed entanglement of cold open-shell polar molecules for universal matchgate quantum computing}

\author{Felipe Herrera}
\address{Department of Chemistry, Purdue University, West Lafayette, IN 47907, USA}
\address{Department of Chemistry and Chemical Biology, Harvard University, 12 Oxford St., Cambridge, MA 02138, USA}

\author{Yudong Cao}
\address{Department of Computer Science, Purdue University, West Lafayette, IN 47907, USA}

\author{Sabre Kais}
\address{Department of Chemistry, Purdue University, West Lafayette, IN 47907, USA}
\address{Qatar Environment and Energy Research Institute, Doha, Qatar}

\author{K. Birgitta Whaley}
\address{Berkeley Quantum Information and Computation Center and Department of Chemistry, University of California, Berkeley, CA 94703, USA}

\date{\today}

\begin{abstract}
Implementing a scalable quantum information processor using polar molecules in optical lattices requires precise control over the long-range dipole-dipole interaction between molecules in selected lattice sites. We present here a scheme using trapped open-shell $^2\Sigma$ polar molecules that allows dipolar exchange processes between nearest and next-nearest neighbors to be controlled to construct a generalized transverse Ising spin Hamiltonian with tunable $XX$, $YY$ and $XY$ couplings in the rotating frame of the driving lasers.  The scheme requires a moderately strong bias magnetic field with near-infrared light to provide local tuning of the qubit energy gap, and mid-infraraed pulses to perform rotational state transfer via stimulated Raman adiabatic passage. No interaction between qubits is present in the absence of the infrared driving. We analyze the fidelity of the resulting two-qubit matchgate, and demonstrate its robustness as a function of the driving parameters. We discuss a realistic application 
of the system for universal matchgate quantum computing in optical lattices.
\end{abstract}

\maketitle

%\tableofcontents

\section{Introduction}

The concept of entanglement has evolved from being regarded as a perplexing and even undesirable consequence of quantum mechanics in the early studies by Schr\"{o}dinger \cite{Schrodinger:1935,Wheeler&Zurek-book} and Einstein \cite{EPR:1935}, to being now widely considered as a fundamental technological resource that can be harnessed in order to perform tasks that exceed the capabilities of classical systems \cite{Horodecki:2009review}. Besides its pioneering applications in secure communication protocols and quantum computing \cite{Nielsen&Chuang-book}, entanglement has also been found to be an important unifying concept in the 
analysis of magnetism \cite{Ghosh:2003,New-Kais1,New-Kais2,Amico:2008review}, electron correlations \cite{New-Kais3,Huang:2005} and quantum phase transitions \cite{Amico:2008review}. Many properties and applications of entanglement have been demonstrated using a variety of physical systems including photons \cite{Aspect:1981,Zhao:2004}, trapped neutral atoms \cite{Mandel:2003,Anderlini:2007,Wilk:2010,Isenhower:2010}, trapped ions \cite{Turchette:1998,Haffner:2005,Blatt:2008}, and hybrid architectures \cite{Blinov:2004,Fasel:2005}. Entanglement has also been shown to persist in macroscopic \cite{Berkley:2003,Yamamoto:2003,Lee:2011} and biological systems \cite{Engel:2007,Sarovar:2010,Zhu:2012}. 

Neutral atomic and molecular ensembles in optical traps are a promising platform for the study of quantum entanglement \cite{Bloch:2005,Bloch:2008}. From a condensed-matter perspective, the large number of trapped particles with highly-tunable interparticle interactions can allow the preparation of novel many-body entangled states using global control fields \cite{Lewenstein:2007,Trefzger:2011,Baranov:2012}. From a quantum computing perspective, optically-trapped neutral particles have long coherence times, and promise the best scalability in comparison with optical, trapped ions, and solid-state architectures \cite{Ladd:2010}. %Moreover, the required coherent manipulation of individual particles using local control fields with sub-micron resolution . Although recent efforts to solve this problem are promising \cite{Weitenberg:2011}, it is important to design control protocols that allow the implementation of universal quantum logic in optical arrays without single-site addressability. 

The long-range character of the interaction between trapped polar molecules \cite{Carr:2009} provides novel mechanisms for entanglement generation and control that are not possible with atoms. 
Arrays of polar molecules can be prepared in optical lattices with full control over the translational, vibrational, rotational and hyperfine degrees of freedom \cite{Ospelkaus:2006,Ni:2008,Ospelkaus:2010-hyperfine,Chotia:2012}. Coherent dipole-exchange interactions between polar molecules in microwave-driven 3D optical lattices \cite{Herrera:2010,Jesus:2010} have recently been observed \cite{Bo:2013}. These experiments pave the way toward the preparation of exotic many-body quantum states with long-range correlations \cite{Baranov:2012}, including topologically-protected dipolar quantum memories \cite{Micheli:2006,Gorshkov:2011prl,Yao:2012,Manmana:2013}. 

Local control of dipolar arrays can also allow the implementation of universal quantum logic within the gate model \cite{Yelin:2009}. 
Two-qubit gates can be implemented spectroscopically using global microwave control pulses, where single-site spectral resolution is provided by an inhomogeneous dc electric field \cite{DeMille:2002,Wei:2011,Zhu:2013, Pellegrini:2011,Mishima:2009}. In this approach the unwanted interactions between qubits can only be suppressed using dynamical decoupling pulses in analogy with NMR architectures \cite{Glaser:2001}. The ability to turn on and off the interaction between selected qubits within a range of sites would greatly simplify the implementation. This approach is taken in Refs. \cite{Yelin:2006,Charron:2007,Kuznetsova:2008} by considering conditional transitions between weakly and strongly-interacting molecular states in dc electric fields, effectively implementing ``switchable'' dipoles. Static electric fields, however, induce dipolar interactions throughout the molecular array that can still introduce undesired two-body phase evolution between qubits that are not participating in the conditional gate. 
In Ref.\ \cite{Kuznetsova:2011}, an atom-molecule hybrid strategy that solves this issue has been proposed.

In this work, we introduce a infrared control scheme to manipulate entanglement between an arbitrary pair of open-shell polar molecules within a range of optical lattice sites. Quantum information is encoded in the spin-rotation degrees of freedom of the molecules in the presence of a bias magnetic field. The controlled two-qubit entangling operation involves the manipulation of local qubit energies using a cw strongly focused near-IR off-resonant laser beam and a single-qubit Raman coherent population transfer step using mid-IR near-resonant laser pulses. Under these conditions, the dipole-dipole interaction is activated for a time sufficient to perform the entangling operation. Reversing the single-qubit control steps suppresses further two-body evolution. Unlike previous proposals for molecular entanglement creation that employ permanent dipoles in dc electric fields, our scheme generates a non-interacting molecular ensemble when the driving fields are not present, or are off-resonant from any 
rovibrational 
transition. We analyze the fidelity of the resulting entangling operation as a function of single-qubit driving parameters. We also discuss how the constructed quantum gates can be used to implement universal matchgate quantum computation in optical lattices.% {\it without} sub-wavelength single-site resolution. 

The remainder of this paper is organized as follows. Section \ref{sec:implementation} introduces the molecular entanglement control scheme and the parameters that characterize the fidelity of the resulting two-qubit operation. In Section \ref{sec:applications} we show how the two-body control scheme implements a set of two-qubit unitaries that can be used to perform universal quantum computing in optical lattices. In Section \ref{sec:discussion} we discuss realistic conditions for physical implementation of the proposed scheme and summarize our findings. 

\section{IR-dressed entanglement generation with two-qubit selectivity}\label{sec:implementation}

We are interested in implementing two-qubit entangling unitaries with trapped polar molecules. Consider an ensemble of $^2\Sigma$ polar diatomic molecules (one unpaired valence electron) in their rovibrational ground state, each individually trapped in a site of an optical lattice in the Mott insulator phase. We assume the molecules are individually trapped in a one-dimensional lattice (along the $x$-axis), which can be prepared from a 3D optical lattice by controlling the trapping wavelength along the $y$ and $z$-axes \cite{Chotia:2012,Bo:2013}. We choose $^2\Sigma$ ground electronic states in for simplicity, but the method described here can be readily generalized for polar molecules with two or more unpaired electrons, including larger polyatomic species. A homogeneous magnetic field allows static control over the valence electron. In addition to the standing-wave weak off-resonant laser that generates the optical trapping potential, we make use of an additional strongly-focused linearly-polarized laser 
beam, far-detuned from any vibronic resonance, that locally enhances the rotational tensor light-shifts only for a subset of lattice sites. As discussed below, static electric fields should be avoided. The collective internal states of the array can be described using the 
Hamiltonian $\mathcal{H} = \sum_i H_i +\sum_{i>j} V_{ij}$, with two-body terms $V_{ij}$ dominated by the dipole-dipole interaction, as discussed below, and one-body terms given by \cite{Carrington:2003}
\begin{equation}
{H}_i = B_e \mathbf{N}_i^2 + \gamma_{\rm sr}\mathbf{N}_i\cdot\mathbf{S}_i + g_S\mu_BB\hat S_{Zi} - U_{\rm LS}(\mathbf{r}_i)\hat C_{2,0}(\theta)\otimes \hat I_S,
 \label{eq:one-body}
\end{equation}
where $B_e$ is the rotational constant, $\mathbf{N}$ is the rotational angular momentum operator, $\mathbf{S}$ the spin angular momentum, $\hat S_Z$ its projection along the quantization axis, and $\hat I_S$ the identity in spin space. The static magnetic field is $\mathbf{B} = B\mathbf{\hat z}$, $g_S\approx 2.0$ is the electron g-factor and $\mu_B$ is the Bohr magneton. For the magnetic field strengths considered in this work, we can ignore the magnetic moment due to the rotation of the nuclei and the hyperfine structure due to the nuclear spin. The last term in Eq.\ (\ref{eq:one-body}) corresponds to the position-dependent tensor light-shift of order $U_{\rm LS}(\mathbf{r}_i)=\Delta\alpha|E_0(\mathbf{r}_i)|^2/4$ for $N\geq 1$, where $\Delta\alpha>0$ is the polarizability anisotropy and $E_0(\mathbf{r}_i)$ is the laser field amplitude seen by the $i$-th molecule, and $N$ is the rotational angular momentum quantum number. The lightshift is due to a strong cw linearly polarized far-detuned near-IR laser that 
is 
several orders of magnitude more intense than the off-resonant trapping light. The spatial dependence of the lightshift results from the ability to focus the strong field so that it interacts with only a subset of the lattice sites when propagating perpendicular to the lattice axis, or with the entire array when propagating along the lattice axis. 
We assume the strong laser polarization is collinear with the magnetic field. The strong laser couples to the rotation via coherent Raman scattering. If we choose $U_{\rm LS}\ll B_e$, the tensor lightshift operator $\hat C_{2,0}(\theta) = (3\cos^2\theta-1)/2$ does not couple rotational states with different values of $N$. The spin-rotation interation $\gamma_{\rm sr}\mathbf{N}\cdot\mathbf{S}$ mixes rotational and spin projections for $N\geq 1$. At the magnetic fields considered here, $\gamma_{\rm sr}/g_S\mu_B B\ll 1$, therefore admixing between the electron spin and the rotational motion of the nuclei is only perturbative. 
The molecular constants $B_e$, $\gamma_{\rm sr}$, and $\Delta\alpha$ depend on the vibrational state of the molecule, although the dependence can be weak for the lowest two vibrational states ($v=0$ and $v=1$) considered in this work \cite{Gonzalez-Ferez:2008}. In Ref.\ \cite{Micheli:2006}, $^2\Sigma$ polar molecules in the regime $\gamma_{\rm sr}/g_S\mu_B B\gg 1 $ were used for the implementation of tunable spin-lattice models. There the dipole-dipole interaction combined with the spin-rotation interaction introduced an effective spin-spin coupling between molecules in the rovibrational manifold $N=0$, via global microwave dressing in a weak dc electric field. In contrast, we use site-local infrared driving within the $N=0$ manifold to induce dipolar exchange processes involving the $N=1$ manifold, as explained below. 

In Eq.\ (\ref{eq:one-body}) we have ignored the quasi-harmonic center-of-mass oscillation of molecules in each lattice site. Such motion corresponds to a phonon bath for the internal state dynamics \cite{Herrera:2010,Herrera:2011}, but the coupling of the collective rotational states with this phonon bath can be made perturbatively weak by increasing the lattice trap frequency. In this regime we can consider molecules to be fixed at the location of the trapping potential minima, and consider the internal state dynamics only.  

% ONE BODY DYNAMICS

We want to implement spin-rotation qubits with a locally-tunable effective bias field. It would be useful to be able to tune the qubit gap to zero in order to suppress the one-qubit phase evolution in the implementation of two-qubit gates, which otherwise would have to be eliminated with additional one-qubit operations. We achieve this by exploiting the unique level structure of open-shell molecules. In Fig. 1 we show the Zeeman spectrum of the lowest two rotational manifolds of a $^2\Sigma$ molecule. For moderately strong magnetic fields (less than 1 Tesla for typical values of $B_e$) the $N=0$ and $N=1$ rotational manifolds cross, as shown in panel 1a. We choose as our computational basis the ground state $\ket{g} = \ket{N=0,M_N=0}\ket{\uparrow}$ and the excited state $\ket{e}\equiv \sqrt{1-a}\ket{N=1,M_N=0}\ket{\downarrow}-\sqrt{a}\ket{N=1,M_N=-1}\ket{\uparrow}$, where $a = \eta^2/2+ \it O\rm(\eta^4)$ and $\eta=\gamma_{\rm sr}/g_{\rm S}\mu_BB_0\ll 1$. The qubit gap $\epsilon \equiv \epsilon_e-\epsilon_g$ 
in this case can be made to vanish when tuning the magnetic field to the location of the energy crossing $B_{\rm cross}$ (see inset 1b). This is a real crossing in the absence of dc electric fields, which would couple the opposite-
parity qubit states to create an avoided crossing. Such parity-breaking fields are always present in experiments, but as long as the interaction energy $U_{\rm dd}=d^2/r_{12}^3$ between adjacent qubits is larger than the linear Stark shift due to stray electric fields, we can consider the crossing to be real and not avoided. For molecules with dipole moments $d\sim 1$ D and lattice site separations $r_{12} \sim 500 $ nm, electric fields of strength $E_{\rm dc}< d/r_{12}^3\sim 1$ mV/cm can be safely ignored. We note that it is possible to suppress stray fields below the mV/cm level in ultracold experiments \cite{Merkt:1998,Sandorfy:1999,Bohlouli:2010}. 

\begin{figure}[t]
\begin{center}
 \includegraphics[width=0.7\textwidth]{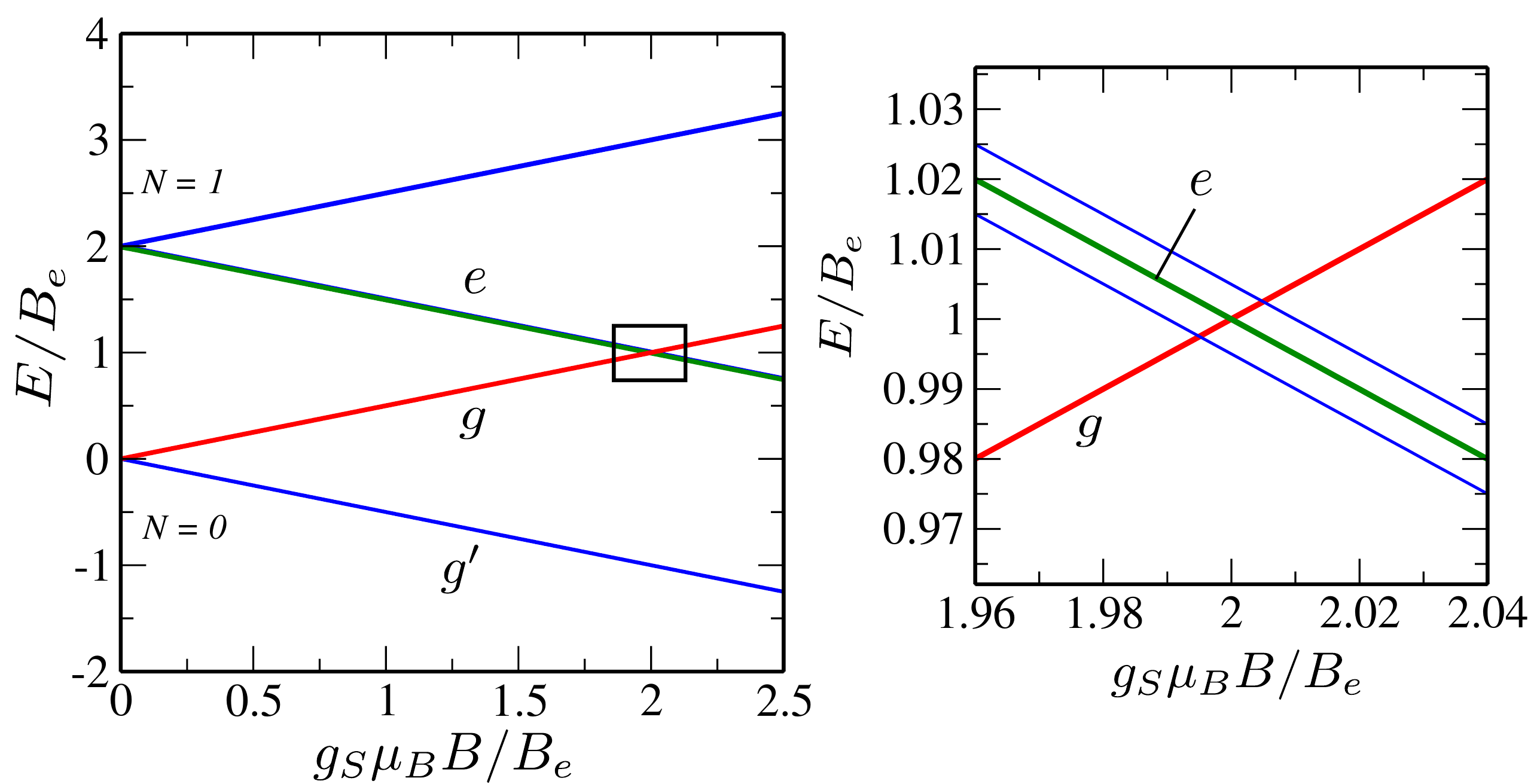}
 \caption{Zeeman spectrum of a $^2\Sigma$ molecule in the ground vibrational state ($v=0$) in a magnetic field $B$. The lowest two rotational manifolds $N=0$ and $N= 1$ are shown. The inset is an expanded view of the squared region near the crossing between the opposite parity qubit states $\ket{g}$ and $\ket{e}$. The separation of $\ket{e}$ from neighbouring excited states is on the order of the spin-rotation constant $\gamma_{\rm sr}/B_e\sim 10^{-2}$. The crossing occurs at the magnetic field $B_{\rm cross} = 2B_e/g_s\mu_B$. $B_e$ is the rotational constant, $g_s$ the electron g-factor and $\mu_B$ the Bohr magneton.}
 \label{fig:zeeman}
 \end{center} 
\end{figure}

\begin{figure}[t]
\begin{center}
 \includegraphics[width=0.6\textwidth]{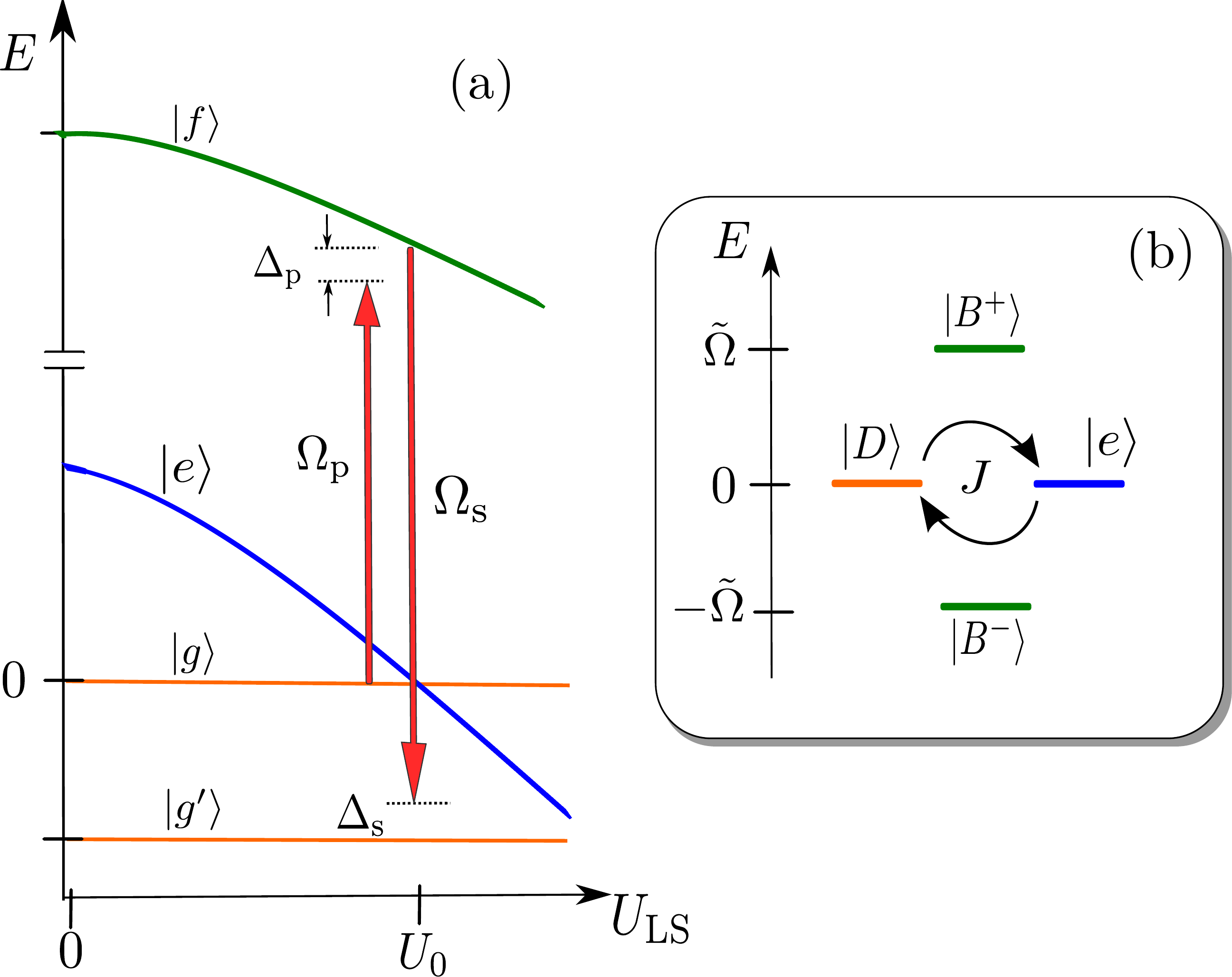}
\end{center} 
 \caption{Level scheme used for infrared dressing. (a) Schematic level diagram for the subspace $\{\ket{g'},\ket{g},\ket{e},\ket{f}\}$ as a function of the tensor lightshift $U_{\rm LS}$. For $U_{LS}=0$, the energies are given by the Zeeman spectrum in Fig. 1. The two ground states $\ket{g}$ and $\ket{g'}$ are coupled to the vibrationally excited state $\ket{f}$. The laser coupling is characterized by the Rabi frequencies $\{\Omega_{\rm p},\Omega_{\rm s}\}$ and detunings $\{\Delta_{\rm p},\Delta_{\rm s}\}$. The state $\ket{e}$ has a tunable gap $\epsilon_e$ from state $\ket{g}$ and is not coupled to $\ket{f}$ by the dressing fields. States $\ket{e}$ and $\ket{g}$ become degenerate when $U_{\rm LS}=U_0$.(b) For $U_{\rm LS}=U_0$, the state $\ket{e}$ is degenerate with the dark-state $\ket{D}=\alpha_1\ket{g}+\alpha_2\ket{g'}$, forming a two-level subspace in which dipole exchange processes occur at the rate $J$. Bright states $\ket{B^{\pm}}$ involving the excited state $\ket{f}$ are separated from the $\{\ket{D}
,\ket{e}\}$ subspace by a gap $\tilde\Omega \gg J$.}
\label{fig:dressing}
\end{figure}

The magnetic field would tune the gap simultaneously for all molecules in the array. The strongly focused near-infrared laser introduced earlier can then be used to manipulate the position of the energy crossing between $\ket{g}$ and $\ket{e}$ locally via the tensor lightshift $U_{LS}(\mathbf{r}_i)$ in the $N=1$ manifold. The energy of the state $\ket{g}$ is unaffected by the tensor lightshift operator $\hat C_{2,0}$ in the regime $U_{\rm LS}\ll B_e$. Note that we are ignoring the state-independent scalar light-shift proportional to the average polarizability $(\alpha_\parallel+2\alpha_\perp)/3$, which lowers the energy of all the rotational states, without affecting the qubit gap $\epsilon_i$. The strong near-infrared laser lowers the energy of the state $\ket{e}$, moving the location of the crossing with $\ket{g}$ to lower magnetic fields. Therefore, local tuning of the gap $\epsilon_i(t)$ can be implemented as follows: ({\it i}) tune the global magnetic field below the energy crossing point and keep 
it fixed 
throughout the experiment; ({\it ii}) change the location of the crossing point quasi-locally (adjacent sites only) by shifting the energy of state $\ket{e}$ using the strongly focused near-IR off-resonant laser; ({\it iii}) refocus the strong laser to manipulate another pair of qubits. For SrF molecules, for example, the energy crossing occurs at the magnetic field $B_{\rm cross} \approx 5376.2$ G for $U_{\rm LS}(\mathbf{r}_i)=0$. If we apply a lower magnetic field $B<B_{\rm cross}$, the gap becomes $\epsilon_e\sim g_S\mu_B|B-B_{\rm cross}|$ for $U_{\rm LS}(\mathbf{r}_i)=0$. A new crossing point is reached when $U_{\rm LS}(\mathbf{r}_i)=U_0$ for $U_0\sim g_S\mu_B|B-B_{\rm cross}|$, making the qubit gap vanish for those sites that are illuminated by the strong near-IR laser. This is illustrated in Fig. \ref{fig:dressing}a, where we set $\epsilon_g=0$. For a typical polarizability anisotropy $\Delta \alpha \sim 100\,a_0^3$ \cite{Meyer:2011}, an off-resonant near IR laser with intensity $I_{\rm LS}\sim 10^2$ 
kW/cm$^2$ is needed to remove a gap $\epsilon_e \sim 1$ MHz. Readily-available cw lasers with 
power $P_{\rm LS}\sim 1$ mW with a beam waist $w_0\sim 1$ $\mu$m can readily achieve these required intensities. 
\\

% TWO BODY DYNAMICS

The two-body dynamics is dominated by the long-range dipole-dipole interaction $\hat V_{ij}$ between molecules in adjacent sites. In the absence of dc electric or near-resonant microwave fields the permanent molecular dipole vanishes, but transition dipole moments between rovibrational states remain finite. We want to exploit this fact to avoid uncontrolled interactions resulting from permanent dipoles. Undesired interactions between molecules need to be compensated using multiple microwave pulses, which increases the complexity of the implementation. Below we show that it is possible to introduce entangling two-body dynamics on demand between neareast neighbours or next-nearest neighbours only, without perturbing the rest of the molecules in the array. The electric dipole-dipole interaction operator can be written as
\begin{equation}
 \hat V_{ij} = U_{\rm dd}(\Theta)\hat D_0^i\hat D_0^j
 \label{eq:two-body}
\end{equation}
where $U_{\rm dd}(\Theta)=({d^2}/{r_{ij}^3})(1-3\cos^2\Theta)$, $r_{ij} = |\mathbf{r}_i-\mathbf{r}_j|$ is the intermolecular distance, $d$ is the body-frame dipole moment of the molecule, $\Theta$ is the angle between the quantization axis and the intermolecular separation vector $\mathbf{r}_{ij}$, and $\hat D_q^i$ is the dimensionless electric dipole operator in spherical coordinates ($q=0,\pm 1$), acting on the $i$-th molecule. Additional terms in $V_{ij}$ involving $\hat D_{\pm 1}$ are strongly suppressed under the dressing conditions described below (see \ref{sec:rf interaction}).  

Even when the qubit gap $\epsilon_e$ can be made to vanish, thus eliminating one-body phase evolution, there is effectively no dipole-dipole interaction in the $\{\ket{g},\ket{e}\}$ subspace because the interaction energy $J_{ij}\propto\bra{e}\hat D_q \ket{g}^2\sim O\rm(\eta^2)$ is only weakly spin-allowed by the spin-rotation interaction ($\eta\ll 1 $). In order to initiate the interaction between molecules, we use stimulated Raman adiabatic passage (STIRAP) \cite{Bergmann:1998} to create a superposition of the state $\ket{g} = \ket{N=0,M_N=0}\ket{\uparrow}$ with its high-field-seeking partner $\ket{g'}=\ket{N=0,M_N=0}\ket{\downarrow}$ within the ground vibrational manifold $v=0$. Specifically, we establish the three-level $\Lambda$ system in Fig. \ref{fig:dressing}a by coupling $\ket{g}$ and $\ket{g'}$ with a common low-field seeking intermediate state $\ket{f}=\sqrt{1-b}\ket{N=1,M_N=-1}\ket{\uparrow}+\sqrt{b}\ket{N=1,M_N=0}\ket{\downarrow}$ in the first excited vibrational state $v=1$, where $b = \eta'^2 
+\it O\rm (\eta'^4)\ll 1$. The dimensionless spin-rotation parameter is $\eta' =\gamma'_{\rm sr}/g_{\rm S}\mu_BB$, where $\gamma_{\rm sr}'$ is the spin-rotation constant in $v=1$. Typically $|\eta-\eta'|/\eta\ll 1$ due to the weak vibrational dependence of the molecular constants for low vibrational quantum numbers. 

The laser coupling scheme is illustrated in Fig. \ref{fig:dressing}a. A left-circularly polarized field, with frequency $\omega_{\rm p}$ in the mid-infrared, couples near-resonantly the states $\ket{g}$ and $\ket{f}$, which have approximately the same spin projection, but opposite parity. A linearly polarized field with frequency $\omega_{\rm s}$ couples the state $\ket{g'}$ with the high-field-seeking component of $\ket{f}$. The driven one-body effective Hamiltonian in the rotating frame becomes
\begin{eqnarray}
 H_i &=& \epsilon_e(t)\ket{e}\bra{e} + \Delta_{\rm p}(t)\ket{f}\bra{f} + [\Delta_{\rm p}(t)-\Delta_{\rm s}(t)]\ket{g'}\bra{g'} + \Omega_{\rm p}(t)\ket{f}\bra{g} + \Omega_{\rm s}(t)\ket{f}\bra{g'} +{\rm H.c.},
 \label{eq:RWA hamiltonian}
\end{eqnarray}
where $\Delta_{\rm p}(t) = \epsilon_f(t) - \omega_{\rm p}$ and $\Delta_{\rm s} = \epsilon_f(t)+\epsilon_g' - \omega_{\rm s}$ are the associated one-photon detunings. $2\Omega_{\rm p}(t) = \bra{f}\mathbf{d}\cdot\mathbf{e}_p\ket{g}E_p(t)$ and $2\Omega_{\rm s}(t) = \bra{f}\mathbf{d}\cdot\mathbf{e}_s\ket{g'}E_s(t)$ are the Rabi frequencies. The transition $\ket{g'}\leftrightarrow\ket{f}$ is only weakly dipole-allowed by the spin-rotation interaction in $\ket{f}$. However, the intensity of the mid-IR driving lasers can be chosen such that the Rabi frequencies $\Omega_{\rm s}(t)$ and $\Omega_{\rm p}(t)$ are of comparable magnitude, within the limits of the rotating-wave approximation. The energies $\epsilon_e(t)$ and $\epsilon_f(t)$ already take into account tensor light-shifts and Zeeman shifts. 
%We assume the dressing field envelopes $E_p(t)$ and $E_s(t)$ vary slowly with respect to the system timescale $1/\epsilon_{f}$.

Under two-photon resonance $\Delta_{\rm p} =\Delta_{\rm s}$, the eigenstates of Eq.\ (\ref{eq:RWA hamiltonian}) include a zero-energy state $\ket{D} = \cos\alpha(t)\ket{g} - \sin\alpha(t)\ket{g'}$ and the states $\ket{B^{\pm}} = (1/\sqrt{2})(\sin\alpha(t)\ket{g}\pm\ket{f}+\cos\alpha(t)\ket{g'})$, with quasi-energies $2\epsilon_\pm = \Delta_{\rm p} \pm \sqrt{\Delta_{\rm p}^2+\Omega_{\rm p}^2+\Omega_{\rm s}^2}$. The mixing angle is $\alpha(t)=\tan^{-1}[\Omega_{\rm p}(t)/\Omega_{\rm s}(t)]$, where for now we take the Rabi frequencies to be real for simplicity. For molecules in the low-field-seeking ground state $\ket{g}$ at some initial time $t_i$, we can write the state vector $\ket{\Psi_i(t_i)}=\ket{D}$ with $\alpha(t_i) = 0$ for $\Omega_{\rm s}\neq 0$. Following the principles of adiabatic passage \cite{Bergmann:1998}, one can prepare the ground-state superposition $\ket{D(t)}$ with $\alpha(t)\neq 0$ by adiabatically tuning the ratio $\Omega_{\rm p}(t)/\Omega_{\rm s}(t)$. Adiabaticity is ensured for driving 
pulses with large area \cite{Bergmann:1998,Vitanov:2001}.

The infrared-dressed two-body interaction in the rotating frame can be obtained by expanding the dipole-dipole interaction operator $\hat V_{ij}$ in the eigenbasis $\{\ket{e},\ket{D},\ket{B^{+}},\ket{B^-}\}^{\otimes 2}$. We now assume one and two-photon resonant driving ($\Delta_{\rm p}=\Delta_{\rm s}=0$) for simplicity, but general expressions for $\Delta_{\rm p}\neq 0$ are straightforward to obtain. Parity conservation of the single-particle bare states restricts the number of non-vanishing interaction matrix elements. We are interested in the two-body dynamics when the energy gap $\epsilon_e\ll |\epsilon_{\pm}|$. In this regime, energetically allowed dipole-dipole transitions are dominated by 
\begin{eqnarray}
\hat V_{ij}  &=& J_{ij}\left\{\ket{e_ie_j}\bra{D_iD_j}+\ket{e_iD_j}\bra{D_ie_j}+{\rm H.c.}\right\}	
\label{eq:dressed two-body}
\end{eqnarray}
where $J_{ij} \equiv (1/3)(d^2/r_{ij}^3)(1-3\cos^2\Theta) (1-\eta^2)(1-\delta^2)$, where $\eta\ll 1$ and $\delta=|\pi/2-\alpha|\ll 1$. Equation (\ref{eq:dressed two-body}) is valid for a small spin admixture of the states $\ket{e}$ and $\ket{f}$ and near complete stimulated Raman adiabatic passage (STIRAP) from $\ket{g}$ to $\ket{g'}$ (see \ref{sec:rf interaction} for details).
Within these constraints, the two-body interaction in the rotating frame of the driving mid-IR fields, together with the one-body term give the effective Hamiltonian 
\begin{eqnarray}
 \mathcal{H} &=& \sum_i \epsilon_i B^{\dagger}_iB_i + \sum_{ij}J_{ij} \left(B^{\dagger}_i+B_i\right)\left(B^{\dagger}_j+B_j \right)\nonumber\\
 &=& \sum_i h_i Z_{i} + \sum_{i}J_{ij} X_iX_{j}
\label{eq:dressed ZXX}
\end{eqnarray}
where $B_i^{\dagger} = \ket{e_i}\bra{D_i}$ creates an excitation in site $i$. In the second line we have used the transformation $B^{\dagger}_i = (X_i+iY_i)/2$ and $B^{\dagger}_iB_i = (1+Z_i)/2$, where $\{X_i,Y_i,Z_i\}$ are Pauli matrices acting on spin $i$, and ignored the constant shift $E = \sum_i\epsilon_i/2$. The ZXX transverse Ising model in Eq.\ (\ref{eq:dressed ZXX}) with effective magnetic field $h_i=\epsilon_e/2$ is widely used to study quantum-phase transitions and non-equilibrium many-body entanglement dynamics \cite{Amico:2008review}.

\begin{figure}[t]
\begin{center}
 \includegraphics[width=0.6\textwidth]{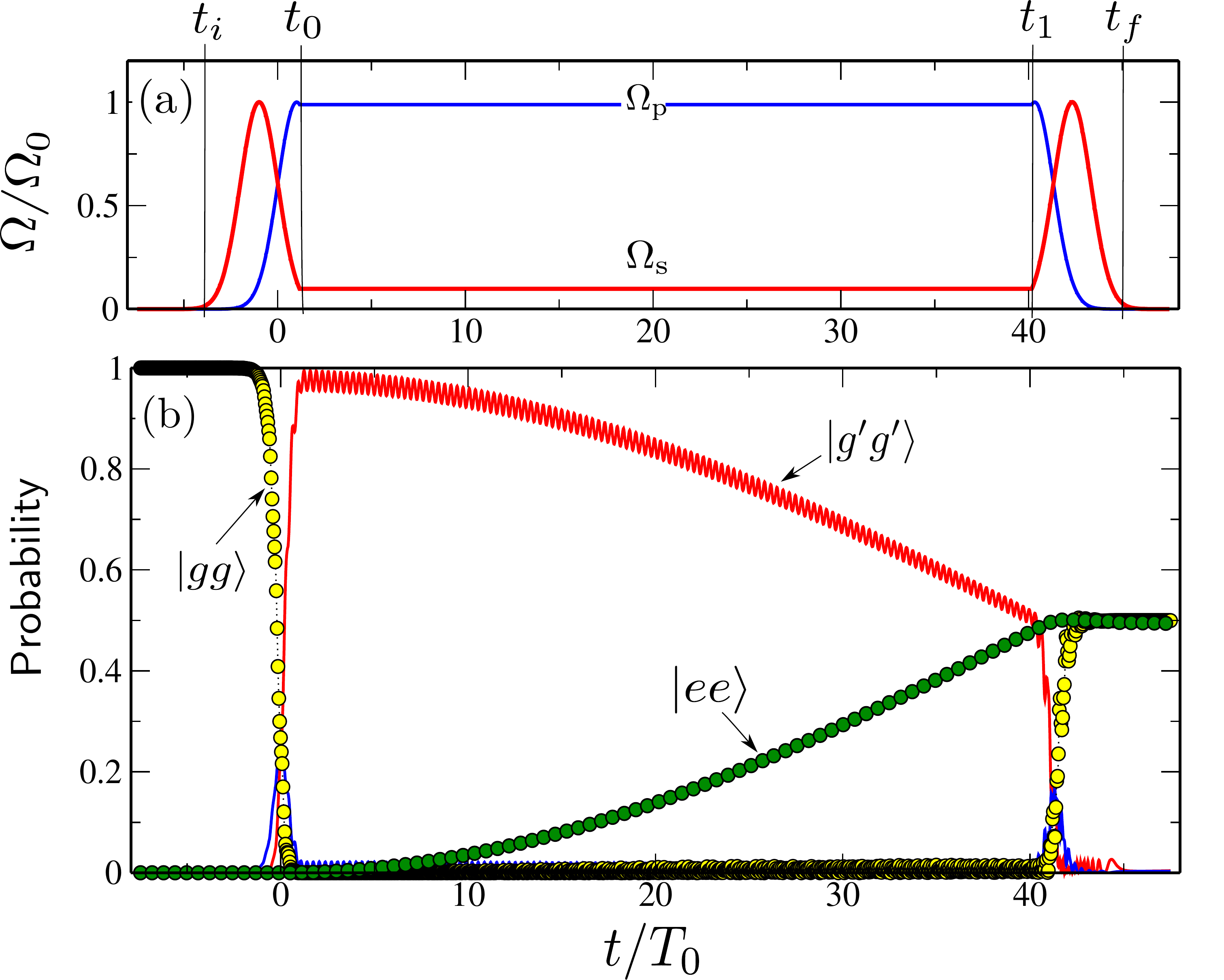}
 \caption{IR-dressed two-qubit gate in the rotating frame. (a) Raman pulse sequence with Stokes pulse $\Omega_{\rm s}(t)$ preceeding the Pump pulse $\Omega_{\rm p}(t)$. At time $t_0$ the mixing angle $\alpha_0$ with $\sin\alpha_0=0.995$ is established. The pulse intensities are then kept constant for a time interval $\tau_e = \pi/4J$. At time $t_1$ the pulse sequence is reversed to return the population to the orginal computational subspace $\{\ket{g},\ket{e}\}^{\otimes 2}$. (b) Two-qubit state evolution in the subspace $\{\ket{g},\ket{e},\ket{f},\ket{g'}\}^{\otimes 2}$, associated with the pulse profile in panel (a), for the input state $\ket{g_1g_2}$. The pulse sequence performs an almost complete population transfer $\ket{g}\rightarrow\ket{g'}$ at time $t_0$. The dipole-dipole interaction between molecules performs the gate $U_{12}(t)={\rm exp}[-iJX_1X_2 t]$, populating the state $\ket{e_1e_2}$. At time $t_1$ the inverse mapping $\ket{g'}\rightarrow\ket{g}$ is performed, leading to the output state $U_{12}
(\tau_{e})\ket{g_1g_2}$. The gate fidelity is $\mathcal{F}_{gg} \equiv |\bra{\Phi_{gg}}U(\tau_{e})\ket{gg}|$, with the ideal output $\ket{\Phi_{gg}}= (\ket{g_1g_2}-i\ket{e_1e_2})/\sqrt{2}$. $\mathcal{F}_{gg}\approx 0.99$ in this example. $t_g=t_f-t_i$ is the total gate time, $J=0.02\,T_0^{-1}$ is the interaction energy and $X_i$ is a Pauli operator. The qubit energy splitting $\epsilon = \epsilon_e-\epsilon_g$ is zero during the gate operation.}
\label{fig:gate}
\end{center} 
 \end{figure}

The ability to engineer the ZXX transverse Ising model in the rotating frame of the driving lasers allows us to implement two-qubit entangling operations as follows:

\begin{enumerate}[ \rm(\it i\rm )]	
 \item Prepare all molecular qubits in their low-field seeking ground states $\ket{g}$. Choose the magnetic field $B<B_{\rm cross}$ below but close to the position of the energy crossing in Fig. 1. This initialization step is done at the begining of the computation and sets the qubit gap $\epsilon_i\neq 0$, for all $i$.
 \item At time $t_i$, eliminate the qubit gap for a chosen pair of molecules using a strongly focused near-IR off-resonant laser field, i.e., $\epsilon_i= \epsilon_{j}=0$, with $|i-j|\leq 2$, 
 \item The mid-IR dressing fields $\Omega_{\rm p}(t)$ and $\Omega_{\rm s}(t)$ resonantly perform the STIRAP mapping $\ket{g}\rightarrow\ket{D}$ only for qubits $i$ and $j$. At time $t_0$, establish the mixing angle $\alpha_0 \equiv\arctan[\Omega_{\rm p}(t_0)/\Omega_{\rm s}(t_0)]= \pi/2-\delta$, with $\delta\ll 1$.
 \item Keep the strong off-resonant laser on for a time $\tau_e\equiv t-t_0= \pi/4J$. In this time interval, the $XX$ interaction implements the maximally entangling gate $U(\tau_e)={\rm e}^{-i(\pi/4)X_iX_{i+1}}$. 
 \item At time $t_1 = t_0+t_e$, reverse the STIRAP pulse sequence to perform the mapping $\ket{D}\rightarrow\ket{g}$ back into the original computational basis. 
 \item At time $t_f$, restore the original qubit gap $\epsilon> 0$ for qubits $i$ and $j$, by turning off the strongly-focused near-IR laser. The total gate time is $t_g = t_f-t_i$.
\end{enumerate}

In Fig. \ref{fig:gate} we illustrate the scheme in steps ({\it i})-({\it vi}) using the input state $\ket{g_ig_j}$ as an example. We find that other inputs give analogous results. The upper panel shows the pulse profile of the driving lasers. For the STIRAP sequence we use delayed Gaussian pulses $\Omega_{\rm p}(t)=\Omega_0{\rm e}^{-(t-\tau_{\rm p})^2/2T_0^2}$ and $\Omega_{\rm s}(t) = \Omega_0{\rm e}^{-(t-\tau_{\rm s})^2/2T_0^2}$, centered at $\tau_p$ and $\tau_s$, respectively (see Fig. \ref{fig:gate}a). We assume the pulses have the same peak Rabi frequency $\Omega_0$ and pulse width $T_0$. We take $\Omega_0 \gg 1/T_0$ to ensure the state $\ket{g}$ evolves into the adiabatic eigenstate $\ket{D}$, suppressing non-adiabatic couplings to the states $\ket{B^\pm}$ \cite{Bergmann:1998,Vitanov:2001}. The state transfer between $\ket{D_iD_j}$ and $\ket{e_ie_j}$ during $\tau_e = \pi/4J$ is shown in Fig 3b, where we use $J = 0.02\,T_0^{-1}$.
At time $t_1$ the molecular pair becomes maximally-entangled in the rotating frame. The STIRAP pulse sequence is then reversed, preserving adiabaticity, in order to return the population to the original computational basis $\{\ket{g},\ket{e}\}$. Non-adiabatic couplings between field-dressed states can move a small amount of population outside the computational basis at the end of the gate sequence. This can affect the overall fidelity of the operation. For the example in Fig. 3, the fidelity is $\mathcal{F}\approx0.996$ both in the rotating-frame (at time $t_1$) and in the computational basis (at time $t_f$). Non-adiabatic couplings can be suppressed by properly designing the laser pulse sequence. The timescale of the complete gate protocol is limited by the dipole-dipole interaction $J$, since short pulses with $T_0\ll \hbar/J$ can always be chosen consistent with the adiabatic restriction by increasing $\Omega_0$, so that the pulse area remains large. We note that a Raman pulse sequence analogous to the 
one in Fig. 3a has been demonstrated using microwave fields to perform gates in trapped ion chains \cite{Timoney:2011}. 
\\

\begin{figure}[t]
\begin{center}
 \includegraphics[width=0.8\textwidth]{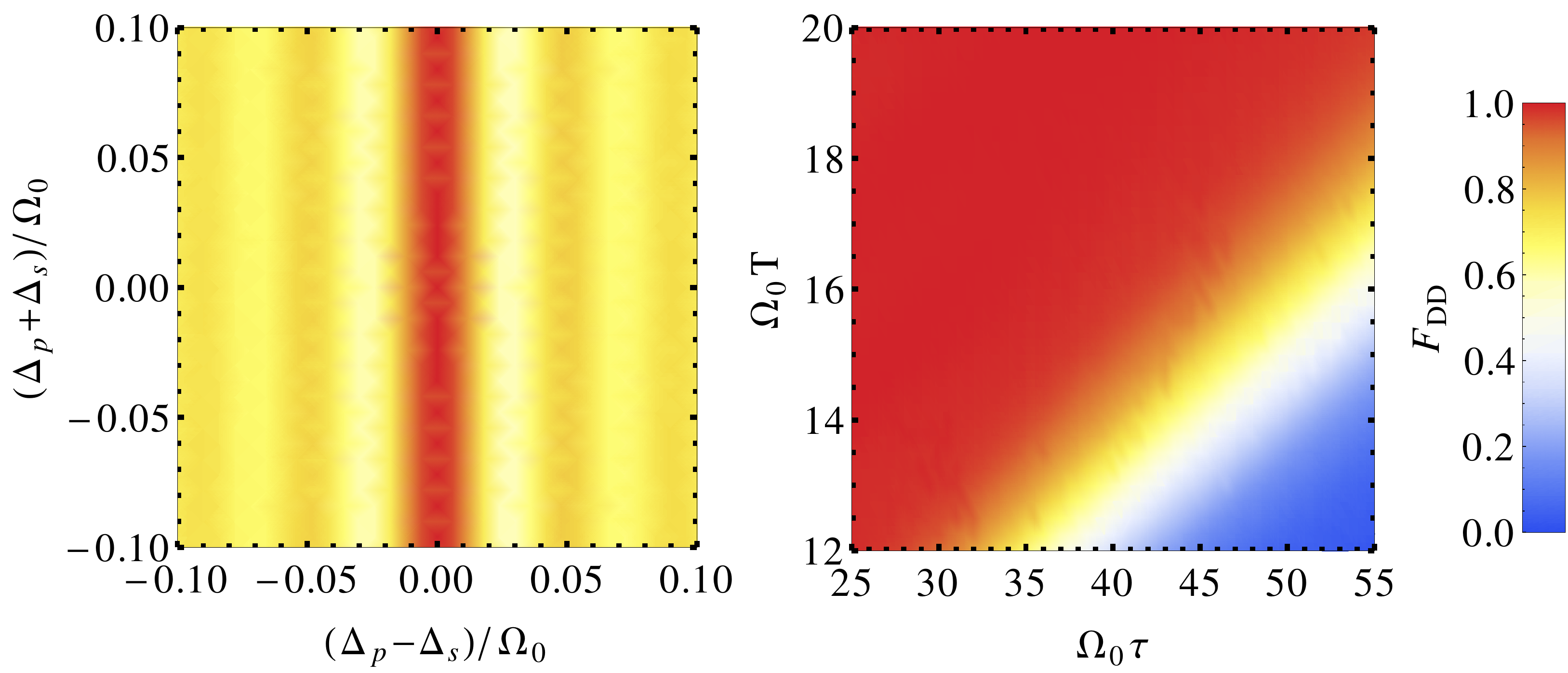}
 \caption{Gate fidelity $\mathcal{F}$ in the rotating frame as a function of the infrared driving pulse parameters. (a) Fidelity $\mathcal{F}_{ab} \equiv |\bra{\Phi_{ab}}U(\tau_{e})\ket{ab}|$ versus two-photon detuning $(\Delta_{\rm p}-\Delta_{\rm s})/\Omega_0$ and $(\Delta_{\rm p}+\Delta_{\rm s})/\Omega_0$, for fixed pulse delay $\tau=40/\Omega_0$ and with $T=20/\Omega_0$ ($\Delta_{\rm p}$ and $\Delta_{\rm s}$ are defined in Fig. 2). The input state is $\ket{ab}=\ket{DD}$, with with $\ket{D}=\cos\theta\ket{g}-\sin\theta\ket{g'}$ and $\sin\theta\approx 0.995$. (b) Fidelity $\mathcal{F}_{DD}$ versus pulse delay $\tau\Omega_0$ and width $T\Omega_0$, under one and two-photon resonance $\Delta_{\rm p} = \Delta_{\rm s} = 0$. The two-qubit gate $U(\tau_e)$ with $\tau_e = \pi/4J$ entangles qubits in the rotating frame. The ideal output state is $\ket{\Phi_{DD}}= (\ket{D_1D_2}-i\ket{e_1e_2})/\sqrt{2}$.  $\Omega_0$ is the peak Rabi frequency of the pulses, which are taken Gaussian with equal width.}
 \label{fig:fidelity}
 \end{center} 
\end{figure}

The robustness of adiabatic population transfer techniques with respect to laser parameters is well-known \cite{Bergmann:1998,Vitanov:2001}. In Fig. \ref{fig:fidelity} we characterize the fidelity of the gate $U(\pi/4J)$ in the rotating frame of the Raman driving with respect to the dimensionless parameters $(\Delta_{\rm p}\Omega_0^{-1},\Delta_{\rm s}\Omega_0^{-1},\tau\Omega_0,T_0\Omega_0)$, where for simplicity we take $\tau_p=-\tau_s=\tau/2$. We define the rotating frame fidelity $\mathcal{F}_{ab}(t)=|\langle \Phi_{ab}|\Psi(t)\rangle|$, where the target state is $\ket{\Phi_{ab}}= U(\pi/4J)\ket{ab}$ with $\{a,b\} = \{D,e\}$, and $\Psi(t)$ is the evolved state in the rotating frame. Figure 4 shows the dependence of the fidelity $\mathcal{F}_{DD}$ with the two-photon detuning $(\Delta_{\rm p}-\Delta_{\rm s})/\Omega_0$ and the sum of detunings $(\Delta_{\rm p}+\Delta_{\rm s})/\Omega_0$. We use the input state $\ket{DD}$, prepared at time $t_0$ by the STIRAP pulses, such that $\sin\alpha_0 \approx 0.995 $. 
Analogous results are obtained for other inputs. 
 For fixed pulsewidth $T_0 = 20/\Omega_0$ and $\tau = 40/\Omega_0$, we obtain $\mathcal{F}_{gg}>99\%$ for a wide range of values inside the band $|(\Delta_{\rm p}-\Delta_{\rm s})|/\Omega_0\leq 0.01$, as shown in Fig. 4a. The two-photon resonance condition is essential to establish the so-called dark state $\ket{D}$ in the rotating-frame, which makes $\mathcal{F}_{ab}$ strongly dependent on the relative detuning of the driving lasers. Note that for two-photon detunings outside the central band with $\mathcal{F}_{DD}>0.9$ in panel \ref{fig:fidelity}a, the fidelity $\mathcal{F}_{DD}$ quickly drops toward the value $\mathcal{F}_{DD}=1/\sqrt{2}$, which corresponds to evolution of the input state $\ket{DD}$ under the identity, i.e., $U(t) = \hat I$, without generation of entanglement.
In Fig. 4b we show the fidelity $\mathcal{F}_{gg}$ under one and two-photon resonance $\Delta_{\rm p}=0$ and $\Delta_{\rm s}=0$, as a function of the dimensionless pulse delay $\tau\Omega_0$ and width $T_0\Omega_0$. Again we find $\mathcal{F}_{gg}>99\%$ for a wide range of parameters. Fidelities below $\mathcal{F}_{DD}=1/\sqrt{2}$ shown in panel 4b correspond to non-overlapping pulses, for which a significant fraction of the evolved state $\ket{\Psi(t)}$ goes outside the $\{\ket{D},\ket{e}\}$ subspace, mostly into the state $\ket{f}=(\ket{B^+}-\ket{B^-})/\sqrt{2}$. For overlapping pulses, both $\tau$ and $T_0$ affect the overlap of the STIRAP pulses, which controls the ratio $\Omega_{\rm p}/\Omega_{\rm s}$. 
The assumed two-photon resonance condition ensures that the state $\ket{D}$ is prepared, but the ratio $\Omega_p(t)/\Omega_s(t)$ determines the mixing angle $\alpha$. For $|\alpha-\pi/2|\sim 1$ dipole-dipole couplings involving $\ket{B^{\pm}}$ are not suppressed. These additional interaction channels move population outside the $\{\ket{D},\ket{e}\}$ qubit subspace, reducing the gate fidelity. The fidelity plots in Fig. \ref{fig:fidelity} also describe the overal gate fidelity in the computational space $\mathcal{F}_{gg}$, provided the state transfer step $\ket{D}\rightarrow\ket{g}$ is efficient.
\\

% HOW TO IMPLEMENT YY + XY + YX INTERACTIONS

Up to now we have restricted our discussion to the implementation of the $ZXX$ Ising model in Eq.\ (\ref{eq:dressed ZXX}) and its associated two-qubit quantum gate. This limitation comes from our choice of vanishing relative phase $\beta = \phi_{\rm p}-\phi_{\rm s}$ between the Raman lasers $\Omega_{\rm p}=|\Omega_{\rm p}|{\rm e}^{i\phi_{\rm p}}$ and $\Omega_{\rm s}=|\Omega_{\rm s}|{\rm e}^{i\phi_{\rm s}}$. However, the relative phase $\beta$ can be controlled experimentally. For $\beta \neq 0$ the dark-state is $\ket{D} = \cos\alpha \ket{g}-{\rm e}^{-i\beta}\sin \alpha\,\ket{g'}$, which in the limit $\alpha=\pi/2-\delta$ gives the electric dipole operator expansion $\hat D_0 = \left\{d'_{eD}\ket{e}\bra{D}+d'_{De}\ket{D}\bra{e}+{\rm H.c}\right\}$ (see Eq. \ref{eq:simplified D0}), with $d_{eD}' = -\sin\alpha \,{\rm e}^{-i\beta} d_{eg'}= d'^{*}_{De}$. The complex phase of $d_{eD}' \equiv \mathcal{A}+i\mathcal{B}$, with $\mathcal{A}=-\sin\alpha\cos\beta d_{eg'}$ and $\mathcal{B} = -\sin\alpha\sin\beta d_{eg'}$, 
is invariant under a global phase 
rotation 
in the subspace $\ket{D}\rightarrow\ket{D}{\rm e}^{i\beta}$ and $\ket{e}\rightarrow\ket{e}{\rm e}^{i\beta}$. The dipole operator can thus be written as $\hat D_{0}= \mathcal{A}\,X -\mathcal{B}\,Y$ and from Eq.\ (\ref{eq:two-body}) we then obtain the expanded interaction term 
\begin{equation}
\hat V_{i,j} = J_{ij}\,X_{i}X_{j} + K_{ij}\,  Y_{i}Y_{j} + L_{ij}\left(X_{i}Y_{j}+Y_{i}X_{j}\right),
\label{eq:YYXY}
\end{equation}
where $J = \mathcal{A}^2 U_{\rm dd}(\Theta)$, $K = \mathcal{B}^2 U_{\rm dd}(\Theta)$, and $L = -\mathcal{A}\mathcal{B} U_{\rm dd}(\Theta)$.

% HOW TO IMPLEMENT ZZ INTERACTION

For some applications it might be interesting to have an interaction term of the form $U_{ij}\ket{e_ie_j}\bra{e_ie_j}$. This type of interaction results from the permanent electric dipole moment in state $\ket{e}$, which we introduce by driving the transition $\ket{e}\rightarrow\ket{e'}$ (not shown in Fig. \ref{fig:dressing}) with a near-resonant cw microwave field characterized by a constant Rabi frequency $\Omega_{\mu}$ and the time-dependent detuning $\Delta_{\mu}(t)$. We choose the state $\ket{e'}=\sqrt{1-c}\ket{v=0;N=2,M_N=0}\ket{\downarrow}+\sqrt{c}\ket{v=0;N=2,M_N = -1}\ket{\uparrow}$, with $c=3\eta^2/2$ and $\eta = \gamma_{\rm sr}/g_s\mu_BB\ll 1$. The microwave frequency is chosen such that it is near resonance with $\Delta E = \epsilon_{e'} - \epsilon_{e}$ only in the presence of the strong near-IR laser field that eliminates the one-body term in Eq.\ (\ref{eq:dressed ZXX}). The microwave field is otherwise far-detuned from any rotational transition with $\Delta_\mu\ll 0$, and only induces a small 
lightshift of order \mbox{$\Omega_\mu^2/|\Delta_\mu|$} to the rotational levels, which can be made much smaller than the spin-rotation constant $\gamma_{\rm sr}$, and therefore negligible, by adjusting the ratio $\Omega_\mu/|\Delta_\mu|$. As the strong near-IR laser changes the detuning $\Delta_{\mu}(t)$, the two-level system $\{\ket{e},\ket{e'}\}$ undergoes chirped adiabatic passage \cite{Vitanov:2001}. This coherent state transfer can be understood using the adiabatic eigenstates of the RWA Hamiltonian in the rotating frame of the microwave field. For an adiabatic change of the detuning satisfying $d\Delta_\mu(t)/dt\ll 2[\Delta_\mu(t)^2+ \Omega_\mu^2]^{3/2}/\Omega_\mu $, the adiabatic state $\ket{e_-(t)} = \cos\theta(t)\ket{e}-\sin\theta(t)\ket{e'}$ with $\tan[2\theta(t)]=\Omega_\mu/\Delta_\mu(t)$, evolves from $\theta(0)=0$ for $\Delta_\mu\rightarrow -\infty$ to $\theta(t)\rightarrow \pi/4$ for as $\Delta_\mu \rightarrow 0$, thus creating a stationary superposition of $\ket{e}$ and $\ket{e'}$ in the 
rotating frame. For $\theta(t)\neq 0$, 
the adiabatic state $\ket{e_-}$ acquires a dipole moment $d_{e_-}\equiv \bra{e_-}\hat D_0\ket{e_-} = -2\cos\theta\sin\theta d_{ee'}$, where $d_{ee'} \approx 2\sqrt{(1-\eta^2)/15}$. The interaction between adjacent permanent dipoles in different lattice sites will then lead to an interaction term of the form 
\begin{eqnarray}
 V_{ij} &=& \sum_{ij} U_{ij} \crea{B}{i}\anni{B}{i}\crea{B}{j}\anni{B}{j} = \frac{1}{4}\sum_{ij} U_{ij}\left(1+2Z_i\right) + \sum_{ij}M_{ij} Z_iZ_j
 \label{eq:ZZ}
\end{eqnarray}
where $U_{ij} = (d^2/r_{ij}^3)(1-3\cos^2\Theta)(d_{e_-})^2$ and $M_{ij} = U_{ij}/4$. Ignoring constant energy shifts, and including the single-qubit terms proportional to $U_{ij}$, Eqs. (\ref{eq:dressed ZXX}), (\ref{eq:YYXY}), and (\ref{eq:ZZ}) may be summarized by the generalized rotating-frame spin Hamiltonian 
\begin{equation}
 \mathcal{H} = \sum_i b_i Z_i +\sum_{ij}J_{ij} X_iX_j + K_{ij}\,  Y_{i}Y_{j} + L_{ij}\left(X_{i}Y_{j}+Y_{i}X_{j}\right)+ M_{ij}Z_iZ_j, 
 \label{eq:dressed XYZ}
\end{equation}
acting on the subspace $\{\ket{D},\ket{e_-}\}^{\otimes 2}$. The parameter space $(J,K,L,M)$ is constrained by $J+K\leq U_{\rm dd}(\Theta)$, $M\leq U_{\rm dd}(\Theta)/4$ and $L^2=JK$, with $U_{\rm dd}(\Theta)=d^2/r_{12}^3(1-3\cos^2\Theta)$. We note that in Ref.\ \cite{Gorshkov:2011}, closed-shell polar molecules in moderately strong dc electric fields were used to implement effective spin-spin couplings of the form in Eq.\ (\ref{eq:dressed XYZ}), plus additional density-dependent terms, via global microwave dressing. In such a system, each parameter can in principle become independently tunable by increasing the number of microwave frequencies used to admix rotational states. In contrast, we use a two-color infrared dressing scheme, which is the simplest scheme that leads to Eq.\ (\ref{eq:YYXY}). Increasing the number of frequencies can allow futher interaction terms, as shown in Eq.\ (\ref{eq:ZZ}) for $ZZ$ couplings. Introducing additional infrared and microwave frequencies can thus decrease the number of 
constraints on the parameters $(J,K,L,M)$. 

The local phase evolution is determined by $b_i = \epsilon_e/2 + \sum_j U_{ij}/2$, which vanishes for $\epsilon_e = -\sum_{j}U_{ij}$. These conditions can be achieved by tuning the magnetic field and the strong off-resonant near-IR laser such that $\epsilon_e <\epsilon_g$. In other words, we can flip the (pseudo)spin qubits by going past a crossing point such as the one illustrated in Fig. 1. Note that in the derivation of Eq.\ (\ref{eq:dressed XYZ}) we have assumed that the Rabi frequency $\Omega_\mu$ is smaller than the coupling constants $J_{ij}$, $K_{ij}$, and $L_{ij}$ so that we can consider the adiabatic state $\ket{e_-}$ to remain quasi-degenerate with the dark state $\ket{D}$.

\section{Universal matchgate quantum computing in optical lattices}
\label{sec:applications}

It is well known that universal quantum computation can be implemented using a maximally-entangling two-qubit gates, most commonly CNOT and CZ, in addition to a minimal set one-qubit rotations \cite{Nielsen&Chuang-book}. In \ref{app:decomposition} we show how to implement CZ and CNOT gates using the two-qubit unitary $U={\rm e}^{-i \mathcal{H}t}$ with $\mathcal{H}$ being the $ZXX$ Ising model in Eq.\ (\ref{eq:dressed ZXX}) for $h_i=0$. Single qubit unitaries in the subspace $\{\ket{g_i},\ket{e_i}\}$ of the $i$-th qubit can be implemented without using mid-IR dressing fields by tuning the local bias field such that $|h_i-h_j|>\varepsilon$ for $j\neq i$, and applying radiofrequency pulses in resonance with $h_i$ that perform arbitrary rotations, in analogy with NMR architectures. The pulse linewidth should satisfy $\gamma_p\ll \varepsilon$. Site resolution of the qubit gap $h_i$ can be achieved using a strongly-focused near-IR laser. Although this this approach has been already implemented for atomic Mott 
insulators \cite{Weitenberg:2011}, we are interested in quantum information processing via two-qubit gates only. 
This goal can be achieved with the matchgate model of quantum computation, which is universal provided that gates may be performed between non-nearest neighbor qubits \cite{Valiant:2002,Terhal:2002,Jozsa:2008}. 

We show here that the physics of interacting molecular transition dipoles can allow for universal matchgate quantum computing in optical lattices. Matchgates $U_{AB}$ \cite{Valiant:2002} are two-qubit unitaries of the form 
\begin{equation}
 U_{AB} = \left(\begin{array}{cccc}
                 a_{11} &0&0&a_{12}\nonumber\\
                 0&b_{11}&b_{12}&0\nonumber\\
                 0 & b_{21} &b_{22}& 0 \nonumber\\
                 a_{21} &0&0&a_{22}
                \end{array}\right),
\label{eq:matchgate}
\end{equation}
where the one-qubit unitaries $A$, with elements $a_{ij}$, and $B$, with elements $b_{ij}$, belong to SU(2) with ${\rm det}(A) = {\rm det}(B)$. 
While quantum computation with matchgates between nearest neighbors in a 1D qubit chain is still efficiently simulable by a classical computer, this is not the case when matchgates between non-nearest neighbor qubits are allowed \cite{Valiant:2002,Terhal:2002,Jozsa:2008,Ramelow:2010}.
%
%It has been proven that an arbitrary quantum computation $U$ composed by a sequence of matchgates acting among nearest-neighbours in a 1D qubit chain, can be efficiently simulated (in polynomial time) by a classical computer \cite{Valiant:2002,Terhal:2002,Jozsa:2008}. However, universal quantum computation using matchgates can still be implemented if two-qubit interactions beyond nearest-neighbours are allowed \cite{Terhal:2002,Jozsa:2008,Ramelow:2010}.
%
%BW 0131 added sentence
We can therefore exploit the long-range nature of the dipole-dipole interaction to realize quantum computations on a 1D chain of dipolar molecules that can not be efficiently simulated by classical means.
In Ref.\ \cite{Jozsa:2008} 
%BW 0131 edited to be more explicit about arbitrary nature of gates from Jozsa and Miyake results
was shown that a universal circuit can be constructed using any set of matchgates acting on nearest-neighbours and next-nearest-neighbour qubits.  In that work,
a demonstration was made for a minimum set of nearest-neighbour matchgates $U_{AB}$ plus SWAP gates: the latter are non-entangling but have the ability to introduce effective long-range interactions between qubits and thus this suffices to ensure universal and non-trivial quantum computation. The price for not using single qubit addressing in the matchgate model of quantum computation is the need to encode logical qubits using two or more physical qubits. This is detrimental for the scalability of matchgate quantum 
computing using currently developed architectures (trapped-ion, solid state, optical), which have a modest number of physical qubits $\mathcal{N}\sim 10$ \cite{Ladd:2010}. Nevertheless, for a 1D optical lattice with $\mathcal{N}\sim 10^2$ physical qubits, encoding a logical qubit using four physical qubits and next-nearest neighbour interactions as in Ref.\ \cite{Jozsa:2008} still gives a processor with a size comparable to 
%BW 0131 inserted 'current' (it is changing rapidly and will be different in a couple of years)
current state-of-the-art trapped ion chains \cite{Haffner:2008,Monroe:2013}. For effective 2D optical arrays with $\mathcal{N}^2$ physical qubits, the computational size would largely exceed currently available implementations
%BW 0131 added clarification
based on other physical systems, even with multi-qubit encoding.

For the polar molecule system described in Sec. \ref{sec:implementation},% work we have demonstrated the ability to engineer the generalized spin Hamiltonian $\mathcal{H}$ in (\ref{eq:dressed XYZ}), using open-shell polar molecules in optical lattices. We note that 
the unitary $U = {\rm e}^{-i\mathcal{H} t}$ is of the form in Eq.\ (\ref{eq:matchgate}), where $\mathcal{H}$ is given by Eq.\ (\ref{eq:dressed XYZ}) with $M_{ij}=0$ \cite{Terhal:2002}. The long-range character of the dipole-dipole interaction between molecules can thus be exploited to ensure universality of the proposed quantum processor under conditions when classical simulation is inefficient \cite{Jozsa:2008}. In Sec. \ref{sec:implementation} we showed that the dipole-dipole coupling is finite only between the those sites for which the mid-IR dressing fields perform the STIRAP transfer that prepares the state $\ket{D}$. We achieved spectral site selectivity by applying a strongly-focused off-resonant near-IR laser to shift the rotational levels. Nearest-neigbour or next-nearest neighbour couplings can thus be implemented by shaping the intensity profile of the near-IR laser with a resolution on the order of the lattice wavelength $\lambda$. Multiple strong beams can also be used. The alternative to direct 
next-nearest-neighbour couplings is to use a SWAP gate to move non-adjacent (
logical) qubits into adjacent locations in 
the lattice. Performing a SWAP gate using $\mathcal{H}$ from Eq.\ (\ref{eq:dressed XYZ}), keeping particles fixed in space, requires single-site addressability \cite{Nielsen&Chuang-book}. Alternatively, physically swapping particles among two lattice sites can effectively implement a SWAP gate, as demonstrated for atoms \cite{Anderlini:2007}. However, this approach requires precise control over the motional state of the particles in the state-dependent lattice potential. The operation involves placing the two particles momentarily in the same lattice site. For molecules, the large number of inelastic collision channels leading to loss of molecules from the trap can make this step challenging to control. Although fermionic suppression of inelastic collisions could be useful to overcome this issue \cite{Shallows:2011,Zirbel:2008}, the required adiabaticity of the lattice spatial 
motion with respect to the lattice trapping period can make the swapping time exceed the millisecond regime \cite{Anderlini:2007}. It might thus be faster and more robust to directly couple next-nearest neighbour qubits by the long-range dipole-dipole interaction  with molecules fixed in space. The associated gate time would be only eight times slower than for adjacent sites. 

\section{Discussion of physical implementation}
\label{sec:discussion}

In section \ref{sec:implementation} we introduced a robust method to engineer the entangling unitary $U(t)$ with two-qubit site resolution. In our analysis of the gate fidelity in Fig. 4, we assumed unitary evolution within the two-particle subspace $\{\ket{g},\ket{f},\ket{g'},\ket{e}\}^{\otimes 2}$. This is only justified if the decoherence rates $\Gamma$ associated with environmental processes are smaller than the entanglement rate $1/t_{\rm e} \propto J_{ij}$. Decoherence times $1/\Gamma\gtrsim  1$ ms have been measured for closed-shell polar molecules in optical lattices \cite{Zhao:2009,Chotia:2012}, resulting from static field fluctuations, incoherent photon scattering off the trapping fields, and motional effects in state-dependent potentials. In the absence of mid-IR and strong near-IR lasers, the system described in this work would be subject to decoherence rates of the similar magnitude as in experiments with close-shell molecules, since the trapping conditions are analogous. The strong far-detuned 
near-IR laser field used to manipulate the qubit gap $\epsilon_e$ (see Fig. 3) can in principle stimulate additional incoherent scattering events that lead to trap loss. The photon scattering rate can be written as $\Gamma_{\rm sc} = {\rm Im}(\alpha)I_{\rm LS}$, where ${\rm Im}(\alpha)$ is the imaginary part of the molecular polarizability and $I_{\rm LS}$ is the near-IR light intensity. The lightshift $U_{\rm LS}(\mathbf{r}_i)$ of the rotational states is proportional to the real part of the polarizability ${\rm Re}(\alpha)$, which for alkali-earth halide $^2\Sigma$ compounds such as SrF is on the order of $10^2$ $a_0^3$ \cite{Meyer:2011}. For light far-detuned from any vibronic resonance the ratio $\rho = {\rm Im}(\alpha)/{\rm Re}(\alpha)$ can be very small. For KRb molecules $\rho = 10^{-7}$ for $\lambda=1064$ nm light \cite{Chotia:2012}, and other polar molecules have similarly low values for $\lambda\sim 1$ $\mu$m \cite{Kotochigova:2006}. Using ${\rm Re}(\alpha) = 100 a_0^3 = 4.6$ HzW$^{-1}$cm$^2$ and $\
rho  = 10^{-7}$ as representative values, the scattering rate for the laser intensities $I_{\rm LS}\sim 10^2$ kW/cm$^2$ considered in Section \ref{sec:implementation} gives the decay time $1/\Gamma_{\rm sc} \approx 20$ s. Molecular dipole moments $d\sim 1$ D and lattice spacings $\lambda/2\sim 500$ nm give dipole-dipole interaction times $1/J\sim 10$ $\mu$s. Two-qubit gates thus occur instantaneously in comparison with the expected photon scattering timescales, introduced by the strong near-IR laser. 

The mid-IR dressing fields could also be a source of trap loss if the intensities are high enough to induce multi-photon vibrational excitation. The peak intensity of the dressing fields is lower-bounded by the approximate adiabaticity condition $\Omega_0T_0\gg 1$, where $T$ is the pulse length and $\Omega_0$ the peak Rabi frequency. Note that since we require the driving fields to remain constant for the duration of the gate, once the adiabatic state $\ket{D}$ is prepared, the parameter $T_0$ more appropriately characterizes the Gaussian turn-on and turn-off times of the beams. We require $T_0$ to be much shorter than $1/J$, so that the preparation of state $\ket{D}$ in Fig. 3 is fast compared with the two-qubit entanglement time $\tau_e = \pi/4J$. For expected interaction times $1/J\sim 10$ $\mu$s, we can choose $T_0\sim 100$ ns, which requires $\Omega_0>$ 10 MHz. Assuming a vibrational transition dipole moment $d= 0.1$ D, intensities $I_0\approx 50$ W/cm$^2$ in the mid-IR spectral region are required to 
ensure adiabaticity. Weakly-allowed 
dipole transitions such as $\ket{g'}\leftrightarrow\ket{f}$ in Fig. 2 are a suppressed by a factor $\eta =\gamma_{\rm sr}/B_e$, at the magnetic fields considered here, with respect to electric dipole-allowed transitions such as $\ket{g}\leftrightarrow\ket{f}$, which have near unity spin overlap. 
However, the required intensities to drive the $\ket{g'}\leftrightarrow\ket{f}$ transition are only a factor $\eta^{-2}$ larger than for $\ket{g'}\leftrightarrow\ket{f}$ in Fig. 3. For SrF molecules $\eta \sim 10^{-2}$ \cite{Meyer:2011}. High intensity mid-IR pulsed and cw laser pulses are commonly used in spectroscopy \cite{Tittel:2003}. Two-photon excitation to higher vibrational states $v>1$ due to the driving fields is strongly suppressed when the associated two-photon detuning is larger than the laser bandwidth.

Since the entanglement creation step involves a strong off-resonant near-IR field, conservative optical dipole forces induced by the light beam can in principle perturb the motion of a molecule in its trapping potential. However, the optical forces of the trapping lasers will dominate the motion of molecules when the spatial intensity inhomogeneity of the strong field is sufficiently small at the position of the molecule. This effect can be estimated using perturbation theory in 1D. A suddenly applied Gaussian beam at $t=0$ creates a lightshift potential of the form $U(x,t) = A_0 {\rm exp}[-x^2/\sigma^2]$ for $t>0$, where $A_0$ is proportional to the polarizability and peak field intensity. A molecule initially trapped in the ground state $\psi_0(x)$ of a static harmonic potential $V_0(x) = m\omega_0^2 x^2/2$, experiences a dipole force that drives transitions to higher motional states, that eventually lead to trap loss. Here $\omega_0$ is the trapping frequency and $m$ is the molecular mass. To lowest order,
 the 
short-time ($\omega_0t\ll 1$) transition probability to the second vibrational mode $\psi_2(x)$ is given by 
$ P_{2\leftarrow 0}(t) =\gamma\left(\gamma - 1\right)^2 A_0^2t^2$
where $\gamma = {\alpha}/({\alpha+\beta})$, $\alpha = m\omega_0/\hbar$ and $\beta = 1/\sigma^2$. Lattice heating is thus suppressed when $\gamma\approx 1$, which requires a beam spatial width $\sigma$ much larger than the trap length $l_0=(\hbar/m\omega_0)^{1/2}$. Otherwise the heating rate is non-perturbative and the strong beam can remove molecules from their traps. This simple estimate shows that it should be possible to increase the lattice frequency $\omega_0$ and shape the intensity profile of the strong near-IR laser field in order to satisfy $l_0/\sigma\ll 1$, reducing the heating rate. Classically, the dipole force from the near-IR beam can be negligible in comparison with the lattice trapping force if the strongest inhomogeneity of the former is pushed to the region in between lattice sites, where no particles are present. Such a beam profile may be produced using perforated screens with slit 
dimensions on the order of the lattice wavelength. This qualitative understanding must be supplemented with more rigorous studies of the heating process, which is subject of future work.
\\

\section{Conclusion}

In this work we have introduced an infrared dressing scheme to implement entangling gates between nearest-neighbour {\it or} next-nearest neighbour open-shell polar molecules in a one-dimensional array. We use $^2\Sigma$ diatomic polar molecules for concreteness, but the scheme is also applicable to diatomic molecules with more than one unpaired valence electron, and polyatomic molecular species. Motivated by recent experimental progress \cite{Weitenberg:2011}, we introduce lattice site selectivity of the infrared dressing scheme using a strongly-focused far-detuned laser that manipulates the energy gap of selected qubits. We choose the molecular qubit states $\ket{g}$ and $\ket{e}$ such that the dipole-dipole interaction between molecules in different sites is negligibly weak in the absence of the infrared driving fields, due to the low spin overlap of the associated transition dipole moments.  
The infrared dressing scheme involves the stimulated Raman adiabatic passage (STIRAP) between the two spin states of the ground rotational manifold $\ket{g}$ and $\ket{g'}$, via an intermediate rotational state in the first vibrationally excited level. Such mid-infrared dressing scheme activates the dipole-dipole interaction between the molecular qubits in selected sites, which we exploit to perform an entangling gate in the rotating frame of the dressing laser fields. Since dc electric fields are not used in the scheme, the molecules remain in a highly-entangled non-interacting state when the dressing fields are absent or are far-detuned from any rovibrational transition. Once the entangling gate is carried out in the rotating frame of the driving fields, the STIRAP step is reversed in order to transfer the entanglement to the original computational basis. We show that the gate time is much faster than the expected decoherence rates, so that the gate fidelity is limited by the efficiency of the STIRAP steps.

We show that the constructed gate belongs to the space of matchgates \cite{Valiant:2002}, which can implement universal quantum computation in one spatial dimension when allowed to act beyond nearest-neighbours \cite{Terhal:2002,Jozsa:2008}. The matchgate model of quantum computation can be particularly useful to implement in optical lattices because it does not require single site addressing. The model requires two-qubit operations only, but encodes one logical qubit in two or more physical qubits.
We then suggest to exploit the long-range character of the dipole-dipole interaction together with the spatial selectivity inherent in our proposed dressing scheme to implement universal matchgate quantum computing, which has yet to be realized experimentally. Such a sytem would allow digital quantum simulations of interacting fermions with Coulomb interactions \cite{Terhal:2002}, which are relevant for quantum chemistry \cite{Kassal:2011}, using polar molecules fixed in the sites of an optical lattice. 

Apart from the possibility of realizing a universal set of matchgates, the extended spin Hamiltonian that we can implement (see Eq.\ (8)) also has an interesting significance from a quantum Hamiltonian complexity perspective \cite{GHL14}. In general, it is known \cite{BL07} that for Hamiltonians of the form in Eq.\ (\ref{eq:dressed XYZ}) with $K_{ij}=L_{ij}=0$ on a 2D square lattice, the worst-case complexity of finding their ground state is {\sc QMA}-complete\footnote{{\sc QMA} is a complexity class that is intended as the quantum analogue of the complexity class {\sc NP}. {\sc QMA}-completeness implies that the problem is hard even on a quantum computer.}. More recent results imply that such Hamiltonians with $L_{ij}=M_{ij}=0$ also have the same property \cite{CBBK13}. Thefore, the realization of the full Hamiltonian with all parameters $(J,K,L,M)$ finite, would represent the simplest controllable quantum system that is hard to simulate even on quantum computers. Moreover, the Hamiltonian $\mathcal{
H}$ in Eq.\ (\ref{eq:dressed XYZ}) is also sufficient for universal adiabatic quantum computation \cite{BL07}. In other words, any quantum circuit could be simulated via adiabatic evolution using the Hamiltonian in Eq.\ (8), by employing appropriate circuit-to-Hamiltonian embedding \cite{BL07}. This suggests that one could in principle use trapped polar molecules for solving {\sc BQP}-complete problems\footnote{{\sc BQP} is complexity class that can be regarded as the quantum analogue of the complexity class {\sc P}.}, which are the hardest problems that quantum computers can solve.

\section{Acknowledgements}

We thank Peter Love for discussions. FH and SK would like to acknowledge the financial support of Purdue Research Foundation. KBW was supported by the National Science Foundation under NSF CHE-1213141 and by DARPA under Award No. 3854-UCB-AFOSR-0041. FH was also supported by DTRA under Award No. HDTRA1-10-1-0046-DOD35CAP.
 %\clearpage 
\appendix
\section{Dipole-dipole interaction in the rotating frame}
\label{sec:rf interaction}

The spherical components of the dimensionless electric dipole tensor $\hat{D}_q$, with $q=-1,0,1$ in the bare basis $\{\ket{e},\ket{g},\ket{f},\ket{g'}\}$ can be decomposed as
\begin{equation}
 \hat{D}_q = d_{g'e}\ket{g'}\bra{e} + d_{g'f}\ket{g'}\bra{f} + d_{gf}\ket{g}\bra{f}+d_{ge}\ket{g}\bra{e}+{\rm H.c.},
 \label{eq:bare dipole component}
\end{equation}
where $d_{g'e} = \bra{g'}\hat{D}_q\ket{e} = d_{eg'}^*$, $d_{g'f} = \bra{g'}\hat{D}_q\ket{f}=d_{fg'}^*$, $d_{gf} = \bra{g}\hat{D}_q\ket{f} = d_{fg}^*$, and and $d_{ge} = \bra{g}\hat{D}_q\ket{e} = d_{eg}^*$. Note that $d_{ge}$ and $d_{g'f}$ are only weakly electric dipole-allowed due to the spin-rotation interaction in excited rotational states. Equation (\ref{eq:bare dipole component}) also holds in the rotating frame of the Raman driving. Transforming to the rotating-frame eigenbasis $\{\ket{e},\ket{D},\ket{B^+},\ket{B_-}\}$ gives the dipole operator components
\begin{eqnarray}
 \hat{D}_0 &=& d'_{eD}\ket{e}\bra{D} + d'_{e+}\ket{e}\bra{B^+} + d'_{e-}\ket{e}\bra{B^-}+ d_{D-}'\ket{D}\bra{B^-}+d_{D+}\ket{D}\bra{B^+}+ {\rm H.c.}\nonumber\\
 &&+ d'_{+-}\left\{\ket{B^+}\bra{B^+}-\ket{B^-}\bra{B^-}\right\} ,
 \label{eq:D0}
\end{eqnarray}
\begin{eqnarray}
 \hat D_1 &=& d_{De}\ket{D}\bra{e} + d_{+e}\ket{B^+}\bra{e} + d_{-e}\ket{B^-}\bra{e}+ d_{D+}\ket{D}\bra{B^+}+d_{D-}\ket{D}\bra{B^-}+ {\rm H.c.}\nonumber\\
 &&+ d_{+-}\left\{\ket{B^+}\bra{B^+}-\ket{B^-}\bra{B^-}+\ket{B^+}\bra{B^-}+\ket{B^-}\bra{B^+}\right\}, 
 \label{eq:D1}
\end{eqnarray}
and $\hat D_{-1} = -\hat D_{1}^{\dagger}$. We define primed dipoles involving $\ket{g'}$ as $d'_{eD} \equiv -(\sin\alpha) d'_{eg'}$, $d'_{e+} \equiv (\cos\alpha) d'_{eg'}/\sqrt{2}=d'_{e-}$, $d'_{D-} = (\sin\alpha) d'_{g'f}/\sqrt{2} = -d'_{D+}$, and $d'_{+-} = (\cos\alpha) d_{g'f}$. The unprimed dipoles involving $\ket{g}$ are $d_{De}\equiv(\cos\alpha) d_{ge}$, $d_{+e} \equiv (\sin\alpha) d_{ge}/\sqrt{2} = d_{-e}$, $d_{D+} \equiv(\cos\alpha) d_{gf}/\sqrt{2}=-d_{D-}$, and $d_{+-} \equiv (\sin\alpha)d_{gf}/2$. Since the states $\ket{B^{\pm}}$ do not have well-defined parity, they acquire permanent dipole moments of size $|d_{+-}+d_{+-}'|$. 

The components $\hat D_{\pm 1}$ do not contribute to the two-body dynamics. The matrix element $d_{ge} = \sqrt{a}\bra{v=0;N=0,M_N=0}\hat D_{1}\ket{v=0;N=1,M_N=-1}$ is suppressed by $a=\eta^2+\it O\rm (\eta^4)$ where $\eta = \gamma_{\rm sr}/g_s\mu_BB_0\ll 1$ at the magnetic fields considered here. Therefore, the matrix elements $d_{De}$, and $d_{\pm e}$ are negligible for all values of the mixing angle $\alpha$. For our choice of excited state $\ket{f}$, the matrix elements $d_{gf}$ cannot be neglected in general. However, since we choose $\delta =\pi/2 - \alpha \ll 1$ the matrix elements $d_{D\pm}\approx \delta\, d_{gf}/\sqrt{2}$ are also suppressed. The remaining dipole moment $d_{+-}\approx (d_{gf}/2)(1-\delta^2/2)$ dominates the expansion of $\hat D_{\pm 1}$, however we can ignore couplings that do not involve $\ket{D}$ and $\ket{e}$ provided the adiabaticity of the one-body state transfer is ensured. 

The component $\hat D_{0}$ can also be simplified. The bare matrix element $d_{g'f}=\sqrt{b}\bra{v=0;N=0M_N=0}\hat D_0\ket{v=1;N=1,M_N=0}$ is suppressed by $b\ll 1$ at the magnetic fields we consider. The matrix elements $d'_{+-}$ and $d'_{D\pm}$ are therefore negligible for all values of $\alpha$ in comparison with those involving $d'_{eg'}$. For $\delta \ll 1$, the term proportional to $d_{e\pm} \approx \delta \, d'_{eg'}/\sqrt{2}$ is also suppressed. The dominant term in the expansion of the dipole operator is thus
\begin{equation}
 \hat D_0 = d'_{eD}\left\{\ket{e}\bra{D} + \ket{D}\bra{e}\right\}.
 \label{eq:simplified D0}
\end{equation}

Using the truncated form of the dimless dipole tensor $\mathbf{D}=\hat D_0 \,\mathbf{e}_0$, the dipole-dipole interaction operator can thus be written as
\begin{equation}
\hat V_{ij} = \frac{d^2}{r_{ij}^3}(1-3\cos^2\Theta) (d_{eD}')^2\left\{\ket{e_ie_j}\bra{D_iD_j} + \ket{D_ie_j}\bra{e_iD_j}+ {\rm H.c.}\right\},
\end{equation}
in the rotating-frame eigenbasis. Note that coupling outside the two-level subspace $\mathcal{S}_1=\left\{\ket{e},\ket{D}\right\}$ is strongly suppressed by the small spin admixture in the bare states $\ket{e}$ and $\ket{f}$, plus the near-complete Raman adiabatic passage from $\ket{g}$ to $\ket{g'}$ ($|\pi/2-\alpha|\ll 1$). Using $d_{eD}'=-(\sin\alpha) d'_{eg'}$ and $d'_{eg'} = \sqrt{(1-\eta^2)}/\sqrt{3}$, we obtain the interaction energy $J_{ij}^{D} \equiv (1/3)(d^2/r_{ij}^3) (1-3\cos^2\Theta) (1-\delta^2)(1-\eta^2)$ in Eq.\ (\ref{eq:dressed two-body}).

\section{Two-qubit entangling gates using single-site resolution: {\tt CZ} and {\tt CNOT}} 
\label{app:decomposition}

The Hamiltonian in Eq.\ (\ref{eq:dressed ZXX}) gives the time evolution operator $U(t)={\rm exp}[-iJ_{12}X_1X_2t]$ when $h_i=0$, $i=1,2$. A controlled-Z gate can be obtained from $U(t)$ via the circuit
%\\*
%With Hadamard transform
\begin{equation}
U_{\sf CZ} = \displaystyle\sqrt{-i}R_z^{(1)}\left(-\frac{\pi}{2}\right)R_z^{(2)}\left(-\frac{\pi}{2}\right)(H_1\otimes{H_2})U\left(\frac{\pi}{4J_{12}}\right)(H_1\otimes{H_2}),
\label{eq:cz}
\end{equation}
where $H_i=(X+Z)\sqrt{2}$ and $R_\nu^{(i)}(\theta)={\rm exp}[-i(\theta/2)\sigma_\nu]$ are the Hadamard and $\nu$-rotation gates acting on qubit $i$, with $\sigma_\nu = \{X,Y,Z\}$. Single-site addressing has been achieved in optical lattices by tuning the qubit gap with sub-micron resolution using an off-resonant optical field and then performing the rotation using long-wavelength radiofrequency or microwave fields \cite{Weitenberg:2011}. 

The {\sf CNOT} gate can then be implemented with the circuit 
\begin{equation}\label{eq:cnot}
U_{\sf CNOT} = \displaystyle iR_z^{(1)}\left({\pi}\right)R_y^{(2)}\left(-\frac{\pi}{2}\right)U_{\sf CZ}R_y^{(2)}\left(\frac{\pi}{2}\right) \\[0.1in]
\end{equation}
with $U_{\sf CZ}$ given by Eq.\ (\ref{eq:cz}).

\clearpage
\bibliography{doubletsigma}
\bibliographystyle{unsrt} 

\end{document}